\documentclass[aps,prl,twocolumn,groupedaddress]{revtex4}

\usepackage{graphicx} 
\usepackage{dcolumn}
\usepackage{bm}

\begin{document}
\title{Pairing mechanism of high-temperature superconductivity: Experimental constraints} 
\author{Guo-meng Zhao$^{1,2,*}$} 
\affiliation{$^{1}$Department of Physics and Astronomy, 
California State University, Los Angeles, CA 90032, USA~\\
$^{2}$Department of Physics, Faculty of Science, Ningbo
University, Ningbo, P. R. China}

\begin{abstract}
Developing a theory of high-temperature superconductivity in copper
oxides is one of the 
outstanding problems in physics. It is a challenge that has defeated theoretical physicists 
for more than twenty years. Attempts to understand this problem
are hindered by the subtle interplay among a few
mechanisms and the presence of several nearly degenerate
and competing phases in these systems. Here we present some crucial
experiments that place essential constraints on the pairing mechanism 
of high-temperature superconductivity. The observed unconventional oxygen-isotope 
effects in cuprates have clearly shown strong electron-phonon interactions and the 
existence of polarons and/or bipolarons.  Angle-resolved photoemission and tunneling spectra
have provided direct evidence for strong coupling to multiple-phonon modes.
In contrast, these spectra do not show strong coupling features expected for magnetic 
resonance modes. Angle-resolved photoemission spectra
and the oxygen-isotope effect on the antiferromagnetic exchange
energy $J$ in undoped parent compounds consistently show that the
polaron binding energy is about 2 eV, which is over one order of magnitude larger than 
$J$ = 0.14 eV.  The normal-state spin-susceptibility data of hole-doped cuprates
indicate that intersite bipolarons are the dominant charge carriers in the 
underdoped region while the component of Fermi-liquid-like polarons is dominant 
in the overdoped region. All the experiments to test the gap or
order-parameter symmetry consistently
demonstrate that the intrinsic gap (pairing) symmetry for the Fermi-liquid-like
component is anisotropic $s$-wave and the order-parameter symmetry of 
the Bose-Einstein condensation of bipolarons is $d$-wave.

\end{abstract}
\maketitle 

\section{BCS theory and the conventional isotope effect on $T_{c}$}

In 1911 curiosity concerning the electrical properties of metals at 
low temperatures led the Dutch physicist, H. K. Onnes and his assistant G. Holst to
discover superconductivity at 4.2 K in mercury \cite{Onnes11}.
This discovery was one of the most 
important experimental findings in low temperature physics.   
Since then tremendous theoretical and experimental 
efforts have been made with the aim of clarifying the microscopic 
mechanism responsible for this novel ground state.

On the long way towards a microscopic understanding of superconductivity, 
the observation of an isotope 
effect on  $T_c$  in 1950 \cite{Maxwell50,Reynolds50} 
gave important clues to the understanding of the microscopic theory of 
superconductivity.  
The presence of an isotope effect thus implies that superconductivity is not 
of purely electronic origin.  In the same year 
H. Fr\"ohlich \cite{Froehlich50} pointed out that the 
electron-phonon interaction gave rise to an indirect 
attractive interaction between electrons, which might be  
responsible for superconductivity. Fr\"ohlich's theory played a decisive role in 
establishing the correct mechanism. Later, in 1956 L. Cooper \cite{Cooper56} 
discovered that electrons with an attractive 
interaction form bound pairs 
(so called Cooper pairs) which lead to superconductivity. 
However, the existence of electron pairs does not necessarily imply a phonon mediated 
pairing. Indeed, Bose condensation 
as considered in 1955 by Schafroth \cite{Schafroth55}, 
is also a possible mechanism for superconductivity, but the 
model was not able to explain the existence of an isotope effect. 
Finally, in 1957, Bardeen, Cooper and Schrieffer \cite{Bardeen57} 
developed the BCS theory that was the first successful microscopic theory of 
superconductivity. 

It is remarkable that the BCS theory shows that
 \begin{equation}
 	k_B T_c = 1.13 \hbar \omega \exp{\left( -\frac{1}{N(0)V}\right) } ~,
 	\label{eq:Tc}
 \end{equation}
where $\omega$ is a typical phonon frequency (e.g., the Debye 
frequency $\omega_{D}$). The electron-phonon coupling constant $N(0)V$ is the product 
of an electron-phonon interaction 
constant $V$ and the electronic density of states at the Fermi surface 
$N(0)$, both of which are assumed to be independent of the ion mass $M$. 
 
This formula implies an isotope-mass dependence of $T_{c}$, with an 
isotope-effect exponent $\alpha$ = $-d\ln T_{c}/d\ln M$ = 1/2, in excellent agreement with the
reported isotope effects  in the non-transition metal superconductors (e.g., Hg, Sn and Pb). 
On the other hand, many superconducting transition metals show a much 
smaller isotope effect (e.g. Zr, Ru: $\alpha = 0 \pm 0.05$), and in some 
materials 
even negative isotope shifts 
have been reported (e.g. U: $\alpha = - 2.2 \pm 0.2$). 
The obvious discrepancy between the 
observed isotope effect and the prediction of the BCS 
theory demonstrates the limitations of the simplified  BCS approach.
Nevertheless, the phonon mediated BCS theory 
was the basis for further and more sophisticated microscopic theories of 
superconductivity. 

Since the original BCS theory of 1957 there has been considerable progress toward 
a deeper
understanding of the electron-phonon interaction. 
In the strong-coupling Eliashberg model \cite{Eliashberg60} 
the electron-phonon coupling constant $\lambda_{ep}$ is related to the spectral function 
$\alpha^2(\omega)F(\omega)$, as defined by the product of the frequency-dependent 
average electron-phonon interaction $\alpha^2(\omega)$ and 
phonon density of states $F(\omega)$.  McMillan \cite{McMillan68} numerically solved
the Eliashberg equations and found an 
expression for $T_c$ taking into account the Coulomb repulsion between electrons and 
the retarded nature of the phonon-induced interaction
 \begin{equation}
 	k_B T_c = \frac{\hbar \omega_D}{1.45}\exp{\left( -\frac{1.04(1+\lambda_{ep})}
 	{\lambda_{ep}-\mu^\ast(1+0.62\lambda_{ep})}\right) },
 	\label{eq:McMillan}
 \end{equation} 
where $\mu^\ast$ is an effective 
Coulomb repulsion. The  BCS result of 
Eq.~(\ref{eq:Tc}) is recovered for $\lambda_{ep} \ll 1$, and in this 
weak-coupling limit
$\lambda_{ep}-\mu^\ast$ substitutes the role of $N(0)V$. 
As a consequence, in strong-coupling superconductors the isotope effect 
exponent 
$\alpha$ is no longer a universal 
value and given by 
\begin{equation}
	\alpha = \frac{1}{2} \left( 1-  
	 \frac{1.04(1+0.62\lambda_{ep})(1+\lambda_{ep})\mu^{\ast 2}}{[\lambda_{ep}-\mu^\ast(1+0.62\lambda_{ep})]^{2}}
	\right) ~.
	\label{eq:ISMcMillan}
\end{equation}

Within the framework of the Eliashberg 
theory a small or even negative isotope exponent is possible.
Therefore, the observation of a small isotope exponent could still be consistent 
with the phonon-mediated pairing mechanism.

\section{Isotope effect on $T_{c}$ in high-$T_{c}$ cuprate
superconductors}

Since the discovery of superconductivity in  
La$_{2-x}$Ba$_{x}$CuO$_4$ by 
Bednorz and M\"uller in 1986 \cite{KAM86}, tremendous efforts have been made 
to clarify the microscopic pairing mechanism 
for high-temeprature superconductivity. 
In some of the cuprates the critical temperatures are far beyond a supposed maximum $T_c \approx 30$~K, 
estimated theoretically on the basis of the conventional phonon-mediated 
mechanism \cite{McMillan68}. However, the theoretical estimate has not
been well justified.

Contrary to low $T_c$ elemental superconductors, the 
cuprates have a complex crystal structure and consist of at least three 
constituents, leading to complicated phononic 
and electronic densities of states. 
The structure of cuprates may be described by the
three main features: i) the CuO$_2$ 
planes, ii) the apical oxygen, iii) the metal-oxygen charge reservoir.
All the high-$T_c$ superconductors have CuO$_2$ planes, but the number of 
planes varies among the different families of cuprates, ranging from a
single-layer up to an infinite-layer structure. 

The investigation of the 
isotope effect in cuprates is certainly much more difficult than 
in elemental superconductors. 
In compound superconductors, such as the cuprates, it is convenient to
define a partial isotope exponent for a constituent with mass $M_i$ 
as
 \begin{equation}
 	\alpha_i = - \frac{\Delta T_c^i}{T_c}\frac{M_i}{\Delta M_i} ~,
 	\label{eq:partial_ois}
 \end{equation}
where the index $i$ denotes a constituent with mass $M_i$ (e.g. 
oxygen). 
The magnitude of $\alpha_i$ 
may provide insight into 
which phonons are important in the occurrence of superconductivity. 
As discussed above, small or even vanishing partial isotope 
exponents do not necessarily imply that the corresponding constituent is not 
important for the occurrence of superconductivity. For a conclusive 
decision on the role of lattice vibrations in 
superconductivity, one should not only 
study the isotope effect on $T_{c}$, but also the isotope effects on some 
other parameters such as the effective supercarrier mass $m^{*}$ and  
the supercarrier density $n_{s}$. The latter effects may offer more 
information about the role of phonons, and will be crucial to the 
understanding of the physics of high-temperature superconductors.

Studies of isotope shifts of $T_{c}$ have been carried out in almost all 
known cuprates. A comprehensive review is given by Franck \cite{Franck94b}. 
Most of the studies reported so far concern the oxygen isotope shift by replacing $^{16}$O with 
$^{18}$O, partly because the experimental procedures are
simple and reliable. Today it is generally accepted that the cuprates doped 
for maximum $T_c$ exhibit a 
small but clearly 
non-vanishing $\alpha_O$. This is true for all the different families 
of doped cuprates. Some of the results are listed in Table.~\ref{tab:OIS}. 
  
\begin{table}[htp]
	\begin{center}
	\begin{tabular}{lcc}
		\hline \hline
		 & & \\
	           & $T_c$ (K) &~~ Oxygen-isotope exponent $\alpha_O$ \\
		\hline
		 & & \\
     La$_{1.85}$Sr$_{0.15}$CuO$_4$ & 35-37  & 0.08-0.13  \\
		
		YBa$_{2}$Cu$_{3}$O$_{6.94}$& 90-92  & 0.02-0.03  \\
		
		YBa$_{2}$Cu$_{4}$O$_{8}$ & 79.5  & 0.06-0.08  \\
			
		Bi$_{2}$Sr$_{2}$CaCu$_{2}$O$_{8+y}$ & 75 & 0.03-0.05   \\
		
		Tl$_{2}$Ca$_2$Ba$_1$Cu$_3$O$_{10}$ & 121 & 0 $\pm 0.12$ \\
		
		 Nd$_{1.85}$Ce$_{0.15}$CuO$_4$ & 24.5 & $<0.05$ \\
		 & & \\
		\hline \hline
	\end{tabular}
	\end{center}
	\caption{Oxygen isotope effect for various cuprates doped for optimum 
	$T_c$ (compiled from Ref.~\protect\cite{Franck94b})
	.}
	\protect\label{tab:OIS}
\end{table} 
Although the different compounds exhibit quite different $T_c$'s, it is 
striking that the oxygen-isotope exponents (OIEs) are similar in magnitude and remain small. 
This suggests that the OIE depends neither on $T_c$ nor on 
specific structural properties such as the number of CuO$_2$ planes and 
the electronic properties of the metal-oxygen buffer layers. The small
OIE observed in the optimally doped cuprates suggest that phonons might 
not be important in high temperature superconductivity. This has led
to many exotic non-phonon mediated pairing mechanisms (e.g., see
Ref.~\cite{Anderson}). 
 
However, the doping dependence of the OIE has been extensively studied 
in different cuprate systems 
\cite{Crawford90,Bornemann92,Franck93,ZhaoLSCO95,Zech95,ZhaoPr,ZhaoJPCM98}
For a particular family of doped cuprates the OIE increases with 
decreasing $T_c$, and can be even 
larger than the BCS prediction \cite{Crawford90}. Such an anomalously large
OIE may 
imply that phonons play an important role in the occurrence of high 
temperature superconductivity in the copper oxides. 
Any correct theories for describing the physics of high-$T_{c}$ 
superconductors must consistently explain both the small OIE in the optimally-doped 
samples, and the anomalously large OIE in the underdoped samples.

The non-vanishing OIEs indicate that 
the oxygen-dominated phonon modes 
are involved in the occurrence of superconductivity in the cuprates.
As we will see below, there are strong coupling features to
multiple phonon modes in tunneling and angle-resolved photoemission
spectra. This implies that there should be isotope shifts related to other atoms. Indeed, large copper-isotope shifts, with $\alpha_{Cu}$ being close to 
$\alpha_O$, have been observed in underdoped  La$_{2-x}$Sr$_{x}$CuO$_4$
(LSCO)  \cite{Franck93} and 
in oxygen-depleted YBa$_{2}$Cu$_{3}$O$_{7-y}$ \cite{ZhaoCu}. This implies that Cu-dominated 
phonon modes also 
play important role in the pairing.

\section{Oxygen-isotope effect on the in-plane supercarrier mass in cuprates}

The conventional phonon-mediated 
superconducting theory is based on the Migdal adiabatic approximation 
in which the phonon-induced electron self-energy is given correctly 
to order $(m/M)^{1/2}$ $\sim$ 10$^{-2}$, where $m$ is the mass of an 
electron. Within this approximation, 
the density of states at the Fermi level  $N(0)$, the electron-phonon coupling 
constant $\lambda_{ep}$, and the effective mass of the supercarriers are all 
independent of the ion-mass $M$. This approximation has been working 
very well in conventional elemental superconductors because the ratio of the sound velocity 
is much smaller than the Fermi velocity. However, since the phase 
velocity of long wave-length optical phonons is large compared to the Fermi 
velocity, it may break down in some compounds with reasonably strong 
electron-phonon interactions \cite{Shif}. The break-down of the Migdal
approximation may lead to the formation of polarons$-$new quasi-particles 
moved together with local lattice distortions. The effective 
mass of polarons $m^{*}$ will depend on $M$. This is because the polaron mass 
$m^{*}= m \exp (\gamma E_{p}/\hbar\omega)$ \cite{alemot,Alex99}, where $m$ 
is the band mass in the absence of the electron-phonon interaction, 
$\gamma$ is a constant, and $\omega$ is a characteristic phonon 
frequency which depends on the masses of ions.  Hence, there should be a 
large isotope effect on the polaronic carrier mass, in contrast to the 
zero isotope effect on the effective carrier mass in ordinary metals. 
Due to the polaronic band narrow effect, the effective density of states at the Fermi 
level  $N(0)$ and the 
effective electron-phonon coupling constant $\lambda_{ep}$
[proportional to $N(0)$] should also depend on the ion mass $M$. This will lead to more  
complicated isotope effect on $T_{c}$. Therefore, negative, vanishing, or small  
isotope exponents $\alpha \ll 1/2$ in optimally doped cuprates do not necessarily imply
that phonons are not important for the pairing mechanism if polarons
are bound into the Cooper pairs.

The total exponent of the isotope effect on the effective carrier mass 
is defined as $\beta = \sum - d\ln m^{*}/d\ln M_{i}$ ($M_{i}$ is the 
mass of the $i$th atom in a unit cell).  For polaronic carriers 
$m^{*}= m \exp (\gamma E_{p}/\hbar\omega)$, this 
definition leads to
\begin{equation}
\beta = -{1\over{2}}\ln(m^{*}/m).
\end{equation}

It is interesting that the simple relation $m^{*}= m \exp (\gamma E_{p}/\hbar\omega)$ 
is even valid in the weak coupling region in the case of the long-range 
Fr\"ohlich electron-phonon interaction \cite{Alex99}.  Then the polaron mass 
enhancement factor $m^{*}/m$ in this case is simply equal to $\exp 
(-2\beta)$. Therefore, if electron-phonon coupling in a solid is strong enough to form 
polarons and/or bipolarons, one should expect a substantial isotope 
effect on the effective mass of carriers.

An important and essential proof for the existence of polarons and/or
bipolarons in cuprates is provided by the observation of the substantial oxygen-isotope effect 
on the penetration depth
\cite{ZhaoYBCO95,ZhaoLSCO95,ZhaoNature97,ZhaoJPCM98,HoferPRL,Zhaoreview1,Zhaoreview2,Zhaoisotope,Keller1,Ro}. 
Zhao {\em et al.} made the first observation of this effect in optimally doped 
YBa$_{2}$Cu$_{3}$O$_{6.93}$ in 1995 \cite{ZhaoYBCO95}.  By precisely 
measuring the diamagnetic signals near $T_{c}$ for the $^{16}$O and $^{18}$O 
samples, the authors were able to deduce the oxygen-isotope effect on the 
penetration depth $\lambda(0)$.  It turns out 
that $\Delta\lambda(0)/\lambda(0)$ = 3.2 $\%$ (Ref.~\cite{ZhaoYBCO95}).

 In fact, for highly anisotropic materials, the observed isotope 
 effect on the angle-averaged $\lambda(0)$ is the same as the isotope 
 effect on the in-plane penetration depth $\lambda_{ab}(0)$. From the 
 magnetic data for YBa$_{2}$Cu$_{3}$O$_{6.93}$, 
 La$_{1.85}$Sr$_{0.15}$CuO$_{4}$, and 
 Bi$_{1.6}$Pb$_{0.4}$Sr$_{2}$Ca$_{2}$Cu$_{3}$O$_{10+y}$, one finds that 
 $\Delta\lambda_{ab} (0)/\lambda_{ab} (0)$ = 3.2$\pm$0.7$\%$ for the
 three optimally doped
cuprates \cite{Zhaoisotope}. Several 
independent  experiments have 
consistently shown that the carrier densities of the two isotope 
samples are the same within 0.0004 per unit cell 
\cite{ZhaoNature97,ZhaoJPCM98,Zhaoreview1,Keller1}. Therefore, the observed oxygen-isotope effect on 
the in-plane 
penetration depth is caused only by the isotope dependence of the 
in-plane supercarrier mass.  Recently, direct measurements of the 
in-plane penetration 
depth by low energy muon-spin-relaxation (LE$\mu$SR) technique 
\cite{Keller1} have confirmed 
the earlier isotope-effect results.  It is found that \cite{Keller1} 
$\Delta\lambda_{ab} (0)/\lambda_{ab} (0)$ = 2.8$\pm$1.0$\%$.  It is 
remarkable that the
isotope effect obtained from the most advanced and expensive 
technology 
(LE$\mu$SR) 
is the same as that deduced from simple magnetic measurements.

\begin{figure}[htb]
\includegraphics[height=6cm]{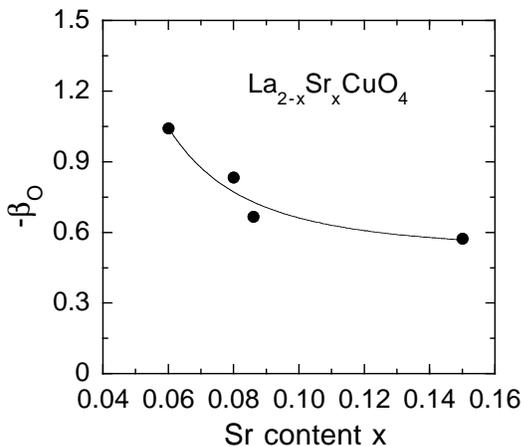}
	\caption[~]{The doping dependence of the exponent 
$\beta_{O}$ of the 
oxygen-isotope effect on the in-plane supercarrier mass in  
La$_{2-x}$Sr$_{x}$CuO$_{4}$. The exponent is defined as $\beta_{O}= 
-d\ln m^{**}_{ab}/d\ln M_{O}$. The data are from 
Ref.~\cite{HoferPRL,Zhaoreview2,Zhaoisotope}}
\end{figure}

Fig.~1 shows the doping dependence of the exponent 
($\beta_{O}$) of the 
oxygen-isotope effect on the in-plane supercarrier mass in  
La$_{2-x}$Sr$_{x}$CuO$_{4}$. Here the exponent is defined as 
$\beta_{O}= 
-d\ln m^{**}_{ab}/d\ln M_{O}$. It is apparent that the exponent 
increases with decreasing doping, in agreement with the fact that 
doping reduces electron-phonon coupling due to screening of charge
carriers.  The large 
oxygen-isotope effect on the in-plane supercarrier 
mass cannot be explained within the conventional phonon-mediated 
pairing mechanism 
where the effective mass of supercarriers is independent of the 
isotope mass \cite{CarbotteRev}. In particular, the substantial 
oxygen-isotope effect 
on $m^{**}_{ab}$ in optimally doped cuprates indicates that the 
polaronic effect is not vanished in the optimal doping regime where the 
BCS-like superconducting transition occurs.  This suggests that 
polaronic carriers may be bound into the Cooper pairs in optimally 
doped and overdoped cuprates. This scenario \cite{alemot,Zhaoisotope} can naturally explain
a small isotope effect on $T_{c}$ in optimally doped cuprates.

\section{Determination of the electron-phonon coupling strength in
undoped parent cuprates}

The observed substantial oxygen-isotope effect on the effective
supercarrier mass suggests that doped holes are strongly
coupled with 
phonons to form polarons and/or bipolarons.   It is natural that 
undoped parent compounds like La$_{2}$CuO$_{4}$ should have stronger electron-phonon
interactions because no doped charge carriers  can screen the electron-phonon interactions. Very strong electron-phonon interactions can reduce electron hopping integral $t$  through the polaronic effect \cite{alemot}, which may lead to the renormalization of  magnetic exchange energy.  Then one might expect that the 
antiferromagnetic (AF) exchange energy  in antiferromagnetic cuprates and the ferromagnetic exchange energy in ferromagnetic manganites should depend on the isotope mass if both systems have very strong electron-phonon coupling.  
Following this simple argument, Zhao and his co-workers initiated 
studies of the 
oxygen-isotope effect on the AF ordering temperature $T_{N}$ in several 
parent cuprates  \cite{ZhaoAF94} and the oxygen-isotope effect on the Curie
temperature in ferromagnetic managnites \cite{ZhaoNature96}. A small oxygen-isotope shift of $T_{N}$ was 
consistently observed in La$_{2}$CuO$_{4}$ \cite{ZhaoAF94} while a
giant oxygen-isotope shift of the Curie temperature was observed in
doped manganites \cite{ZhaoNature96}. The large difference in the
isotope effects on the magnetic ordering temperatures appears
contradictory since both materials should have similar electron-phonon 
coupling strengths. Here we will show that the small isotope shift of $T_{N}$
is also consistent with a large polaron binding energy due to the fact that the electron-phonon interaction renormalizes the antiferromagnetic exchange energy $J$ through the forth order process \cite{Eremin}.

From the oxygen-isotope shift of $T_{N}$, one can determine the oxygen-isotope 
effect on the antiferromagnetic exchange energy $J$, that is, $\Delta J/J$ $\simeq -0.6\%$
(Ref.~\cite{ZhaoAF94}).
A slightly larger oxygen-isotope effect on $J$ ($\Delta J/J$ $\simeq$
$-$0.96$\%$) has also been extracted from
the mid-infrared two-magnon absorption spectra of the oxygen-isotope exchanged
YBa$_{2}$Cu$_{3}$O$_{6.0}$ (Ref.~\cite{ZhaoAF07}). This unconventional isotope effect on $J$ can be explained in terms of
strong electron-phonon coupling and polaronic effect. Eremin {\em et al.} \cite{Eremin} 
have considered strong 
electron-phonon coupling within a three-band Hubbard model appropriate for
the charge-transfer insulators.  They show that the antiferromagnetic exchange energy $J$ depends on the polaron 
binding energy $E_{p}^{O}$ due to oxygen vibrations, on the polaron 
binding energy $E_{p}^{Cu}$ due to copper vibrations, and on their 
respective vibration frequencies $\omega_{O}$ and $\omega_{Cu}$. At 
low temperatures, $J$ is given by \cite{Eremin}
\begin{equation}\label{J}
J = J_{\circ}(1 + 
\frac{3E_{p}^{O}\hbar\omega_{O}}{\Delta_{pd}^{2}}+\frac{3E_{p}^{Cu}\hbar\omega_{Cu}}{\Delta_{pd}^{2}}).
\end{equation}

Here $\Delta_{pd}$ is the charge-transfer gap, which is measured to 
be about 1.5 eV in undoped cuprates. From the above equation, one can 
clearly see that there should be an observable oxygen-isotope effect on $J$ 
if the polaron binding energy is significant. The oxygen-isotope effect on $J$ 
can be readily deduced from Eq.~\ref{J}:
\begin{equation}\label{OIEJ}
\frac{\Delta J}{J} = 
(\frac{3E_{p}^{O}\hbar\omega_{O}}{\Delta_{pd}^{2}})(\frac{\Delta\omega_{O}}{\omega_{O}}).
\end{equation}

The charge-transfer 
gaps $\Delta_{pd}$ have been measured for both La$_{2}$CuO$_{4}$ and 
YBa$_{2}$Cu$_{3}$O$_{6}$ systems \cite{Cooper}, that is, 
$\Delta_{pd}$ = 1.81 eV for La$_{2}$CuO$_{4}$ and $\Delta_{pd}$ = 1.60 
eV for YBa$_{2}$Cu$_{3}$O$_{6}$. If we take  $\hbar\omega_{O}$ = 
0.075 eV and substitute the above parameters into Eq.~\ref{OIEJ}, we obtain 
$E_{p}^{O}$ = 1.46 eV for La$_{2}$CuO$_{4}$ and $E_{p}^{O}$ = 1.82 eV 
for YBa$_{2}$Cu$_{3}$O$_{6}$. Since the copper-related phonon modes have much lower 
energies than the oxygen-related modes, the copper-related phonon 
modes may contribute little to the polaron formation, that is, 
$E_{p}^{Cu}$ $<<$ 1. Therefore, the total polaron binding energy 
$E_{p}$ should be about 1.5 eV and 1.9 eV for La$_{2}$CuO$_{4}$ 
and YBa$_{2}$Cu$_{3}$O$_{6}$, respectively.

Angle-resolved photoemission spectroscopy (ARPES) data of 
undoped La$_{2}$CuO$_{4}$ have been explained in terms of polaronic 
coupling between phonons and charge carriers \cite{Ros}.  From the 
width of the phonon side band in the ARPES spectra, the authors find 
the polaron binding energy to be about 1.92 eV, in good agreement with 
their theoretical calculation based on a shell model \cite{Ros}.  On the other 
hand, the observed binding energy of the side band should be 
consistent with a polaron binding energy of about 1.2 eV (Ref.~\cite{Ros}).  
This should be the lower limit because the binding energy of the side band 
decreases rapidly with doping and because the sample may be lightly doped 
\cite{Ros}.  Therefore, the ARPES data suggest that 1.2 eV $<$ $E_{p}$ 
$<$ 1.9 eV, which is in quantitative agreement with the value ($E_{p}$ $>$ 1.5 eV) deduced 
from the isotope effect on the exchange energy.

The parameter-free estimate of the polaron 
binding 
energy due to the long-range Fr\"ohlich-type electron-phonon 
interaction 
has been made for many oxides including cuprates and manganites 
\cite{AlexJPCM99}. The 
polaron binding energy due to the long-range Fr\"ohlich-type electron-phonon 
interaction is estimated to be about 0.65 eV for La$_{2}$CuO$_{4}$ 
(Ref.~\cite{AlexJPCM99}). The polaron 
binding energy due to the $Q_{1}$-type Jahn-Teller distortion is about 
1.2 eV for La$_{2}$CuO$_{4}$ (Ref.\cite{Kam}). The total polaron binding energy $E_{p}$ should 
be at least 1.85 eV, in excellent agreement with the value ( 1.5 eV $<$ $E_{p}$
$<$ 1.9 eV) deduced independently from the oxygen-isotope effect on $J$ and ARPES data.

If there are very small amounts of charged carriers in these nearly 
undoped compounds, the optical conductivity due to the polaronic 
charge carriers will show a broad peak at 
$E_{m}$ = 2$\gamma E_{p}$ (Ref.\cite{AlexJPCM99}), where 
$\gamma$ is  0.2$-$0.3 for layered cuprates \cite{AlexJPCM99}.  There appears to exist the 
third broad peak at 0.7-0.8 eV in the optical conductivity of 
Sr$_{2}$CuO$_{2}$Cl$_{2}$ (Ref.~\cite{Choi}),  La$_{2}$CuO$_{4}$ (Ref.~\cite{Perkins}), and 
YBa$_{2}$Cu$_{3}$O$_{6}$ (Ref.~\cite{Grun}). This peak should be caused by the polaronic 
effect because the energy scale for the peak is similar to that 
predicted from the polaron theory assuming $\gamma$ $\sim$ 0.2. Hole doping will reduce 
the value of $E_{p}$ and thus $E_{m}$ due to screening of charged 
carriers. Indeed, $E_{m}$ was found to be  about 0.6 eV for 
La$_{1.98}$Sr$_{0.02}$CuO$_{4}$ and  0.44 eV 
for La$_{1.94}$Sr$_{0.06}$CuO$_{4}$ (Ref.~\cite{Bi}).

\section{Strong coupling to multiple-phonon modes in angle-resolved photoemission
spectroscopy and tunneling spectra}

It is known that the electron self-energy
renormalization effect in the form of a ``kink'' in the band
dispersion can
reveal coupling of electrons with phonon modes. The ``kink'' feature 
at an energy of about 70 meV has been seen in the band dispersion of 
various 
cuprate superconductors along the diagonal (``nodal'') direction 
\cite{Lanzara01}. The energy scale of about 70 meV appears to provide evidence 
for strong coupling between electrons and the 70 meV Cu-O 
half-breathing mode observed by neutron scattering \cite{McQueeney}. From 
the measured dispersion, one can extract the real part of electron 
self-energy that contains information about coupling of electrons 
with 
collective boson modes. The fine coupling structures can be 
revealed in the high-resolution ARPES data. Very recently, such 
fine coupling structures have been clearly seen in the raw data of 
electron self-energy 
of deeply underdoped La$_{2-x}$Sr$_{x}$CuO$_{4}$ along the diagonal 
direction \cite{Zhou04}. Using the maximum entropy method (MEM) 
procedure, they are 
able to extract the electron-phonon spectral density 
$\alpha^{2}F(\omega)$ that contains coupling features at 27 meV, 45 
meV, 61 meV and 75 meV. These ARPES data and exclusive data 
analysis \cite{Zhou04} clearly indicate that the phonons rather than 
the magnetic collective modes are responsible for the electron self-energy effect.

In fact, the fine coupling structures also appear in the earlier 
high-resolution ARPES data of a slightly overdoped BSCCO with $T_{c}$ 
= 91 K (Ref.~\cite{Johnson}). Zhao \cite{ZhaoSM07} reanalyzed the ARPES data and 
also showed strong coupling features at 21.4 meV, 38.3 meV, 44.4 meV, 60.0 meV, and 76.2 meV, 
It is remarkable that the peak
positions at 44.4 meV, 60.0 meV, and 76.2 meV for BSCCO match
precisely with the peak positions at 45 meV, 60 meV, and 76 meV for 
underdoped La$_{2-x}$Sr$_{x}$CuO$_{4}$ (Ref.~\cite{Zhou04}). This consistency suggests that
the fine structures seen in the high-resolution ARPES data are indeed 
related to strong electron-phonon coupling to multiple phonon modes.

Strong electron-phonon coupling to multiple phonon modes have been also  seen in several high-quality single-particle tunneling spectra
\cite{Shim,Gonnelli,Zhaopairing07,Boz08}. For conventional 
superconductors, 
strong electron-phonon coupling features clearly show up in 
single-particle tunneling spectra \cite{MR,CarbotteRev,MRbook}. The energies  of the 
phonon modes coupled to electrons can be precisely determined from tunneling 
spectra. More specifically, the energy positions of the peaks in 
$-$d$^{2}I$/d$V^{2}$, 
measured from the isotropic $s$-wave superconducting gap $\Delta$, 
correspond to those of the peaks in the electron-phonon spectral function 
$\alpha{^2}(\omega)$$F(\omega)$ (Refs.~\cite{MR,CarbotteRev,MRbook}). Therefore, if 
the coupling strength $\alpha^{2}(\omega)$ changes smoothly with
$\omega$, the fine structures in the phonon density of states $F(\omega)$ 
will have a one-to-one correspondence to those in 
$-$d$^{2}I$/d$V^{2}$.  Such a conventional approach to identify the 
electron-boson coupling features would have been extensively applied 
to high-$T_{c}$ superconductors if the superconducting gap were 
isotropic.  Since the superconducting gap is highly anisotropic in 
high-$T_{c}$ superconductors, it is difficult to reliably determine 
the energies of bosonic modes if tunneling current is not 
directional.  This may explain why the electron-boson coupling structures 
extracted from earlier tunneling spectra are not reproducible among 
different groups \cite{Ved,Shim,Gonnelli}.  

Zhao \cite{Zhaopairing07} analyzed high-quality tunneling spectra of YBa$_{2}$Cu$_{3}$O$_{7-\delta}$ 
and Bi$_{2}$Sr$_{2}$CaCu$_{2}$O$_{8+\delta}$. The spectra of the second 
derivative of tunneling current 
d$^{2}I$/d$V^{2}$ in both compounds show clear dip and peak features due to strong coupling to the 
bosonic modes mediating electron pairing. The energy positions of 
nearly all the peaks in $-$d$^{2}I$/d$V^{2}$-like spectra match 
precisely with those in the phonon density of states obtained by 
inelastic neutron scattering. Similar results have also been reported 
in a La$_{1.84}$Sr$_{0.16}$CuO$_{4}$ thin film \cite{Boz08}. These results 
consistently demonstrate that 
the bosonic modes mediating the electron pairing are 
phonons and that the large attractive interaction should arise primarily from strong 
coupling to multiple phonon modes. 

\begin{figure}[htb]
	 \includegraphics[height=6.2cm]{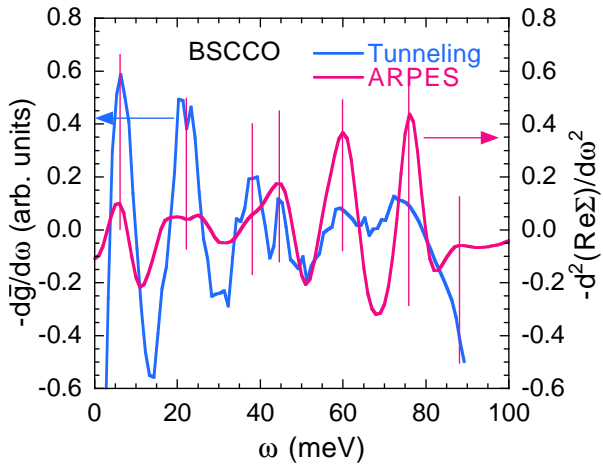}
	\caption[~]{The 
$-$d$^{2}$(Re$\Sigma$)/d$\omega^{2}$ spectrum ($\Sigma$ is electron
self-energy)  and the $-$d$\bar{g}$/d$\omega$ tunneling 
spectrum of BSCCO crystal. After Ref.~\cite{ZhaoSM07}. }
\end{figure}

In Fig.~2, we compare the strong coupling features revealed in the
tunneling spectrum and in the electron self-energy spectrum of BSCCO.  The spectrum of the 
second derivative of the real part of electron self-energy ($\Sigma$) 
is reproduced from Ref.~\cite{ZhaoSM07}. It is striking that the peak features 
in ($-$d$^{2}$(Re$\Sigma$)/d$\omega^{2}$) 
match very well with those in ($-\Delta$(d$\bar{g}$/d$\omega$), where $\bar{g}$
is the renormalized tunneling conductance proportional to the first
derivative of tunneling current \cite{Zhaopairing07}. This excellent agreement 
between the tunneling and self-energy spectra further demonstrates that the 
observed strong-coupling features in both spectra are intrinsic.

On the other hand, the authors of Ref.~\cite{Nie} mistakenly
assigned the energy
(10.5 meV) of a peak position  in $+$d$^{2}I$/d$V^{2}$ spectra of 
an electron-doped Pr$_{0.88}$LaCe$_{0.12}$CuO$_{4}$ to the energy of a bosonic mode. Since this mistakenly
assigned mode
energy (10.5 meV) is very close to the energy (9.5-11 meV) of the magnetic
resonance-like mode measured by inelastic neutron scattering
\cite{Wilson,JZhao}, the authors \cite{Nie} conclude that the 
magnetic resonance mode mediates electron pairing in electron-doped cuprates. Zhao 
\cite{ZhaoPRL09} reanalyzed the tunneling data and found strong coupling to a 
bosonic mode at about 16 meV, in quantitative agreement with early tunneling
spectra of Nd$_{1.85}$Ce$_{0.15}$CuO$_{4}$ (Ref.~\cite{Huang}). The
correctly assigned mode energy of 
about 16 meV rules out its
 connection to the magnetic resonance mode which has energy of 9.5 meV
 in NCCO (Ref.~\cite{JZhao}) and 11 meV in PLCCO (Ref.~\cite{Wilson}).
 
The pairing mechanism based on strong coupling to a 
magnetic resonance mode predicts a pronounced dip feature 
at the energy of the magnetic resonance mode in tunneling conductance spectra 
and angle-resolved photoemission spectra \cite{Abanov}. Some tunneling
spectra of optimally doped BSCCO appear to support this prediction.
For example, the energy of the dip feature in the tunneling spectrum
of BSCCO is found to be about 45 meV
(Ref.~\cite{Gonnelli}), close to the
magnetic resonance energy of about 43 meV (Ref.~\cite{Fong}). However, 
the inversion of the same tunneling spectrum \cite{Gonnelli} yields a very strong coupling feature at
about 20 meV, in quantitative agreement with the mode energies of
several overdoped BSCCO crystals inferred from the electron
self-energies along the antinodal direction \cite{Zhaopairing05}. 
Furthermore, this theoretical
prediction also contradicts more recent atomic-resolution 
tunneling spectra in various BSCCO crystals. The observed dip features in 
these crystals center around 26 meV
and are nearly independent of doping or $T_{c}$ \cite{Lee}. 
Since the energy of the magnetic resonance mode is found to be proportional to
$T_{c}$ (Ref.~\cite{He}), the fact that 
the dip features are nearly independent of doping suggests that they
are not caused by strong coupling to the magnetic resonance modes. 
Furthermore, the energy (26 meV) of the dip feature in optimally doped BSCCO is too low 
compared with the energy (43 meV) of the magnetic resonance mode and the energy (35-40 meV) of the  out-of-phase bond buckling oxygen modes \cite{John2010}.

Further evidence for no magnetic coupling is magneto-optical experiments on 
various cuprates \cite{YLee,Dord}. The data have shown  that the
electron-boson spectral function  is independent
of magnetic field \cite{YLee}, in contradiction with the theoretical
prediction based on the magnetic pairing mechanism (see
Fig.~9 of Ref.~\cite{YLee}).

\section{Co-existence of bipolarons (formed between apical
and in-plane oxygen holes) with in-plane oxygen hole polarons}

It is known that undoped parent cuprates are charge-transfer
insulators, so doped holes should mainly reside on the oxygen orbitals.  
Bulk-sensitive x-ray-absorption experiment on
(Y$_{1-x}$Ca$_{x}$)Ba$_{2}$Cu$_{3}$O$_{7-y}$ confirms that doped holes
mainly reside on both in-plane oxygen and apical oxygen orbitals
\cite{Merz}. What is
striking is that doped holes via oxygen
doping in the CuO chains are almost
equally distributed to the apical and in-plane oxygen orbitals \cite{Merz}. 
The Ca substitution for Y yields hole doping into the in-plane oxygen 
orbitals only and no superconductivity is seen in a compound with
10$\%$ in-plane oxygen holes and negligible apical oxygen holes. This 
remarkable experiment clearly demonstrates that the apical oxygen
holes are crucial to high-temperature superconductivity. The data
provide strong support for the theory of bipolaronic superconductivity
proposed by Alexandrov and Mott \cite{alemot}. The theory is based on 
mobile intersite bipolarons that are formed between apical and in-plane oxygen holes.
Different types of
bipolarons in La$_{2}$CuO$_{4}$ were investigated by 
Catlow {\em et al.} \cite{Cat} with computer simulation techniques based on the
minimization of the ground-state energy without the hopping term. An intersite 
bipolaron formed between 
an in-plane and an apical oxygen-hole polaron (denoted by AP
bipolaron) was
found energetically favorable with the binding energy of 0.14 eV. An intersite 
bipolaron
formed between two in-plane oxygen hole polarons (denoted by PP bipolaron) is also possible 
but with a much
smaller binding energy (0.06 eV). Both electron-phonon interactions
and bipolaron binding energies will decrease when the hoping term
is turned on. Doping will further reduce the bipolaron binding energy due to the screening of charge carriers. Therefore, it is possible that  only AP bipolarons are stable in doped cuprates. For YBCO, about half of the
inplane oxygen holes will be combined with the apical oxygen holes to 
form intersite bipolarons while the remaining in-plane oxygen holes
are mixed with in-plane Cu holes and may be bound into $k$-space pairs below $T_{c}$. 
For LSCO, the number 
of the apical oxygen holes might be quite close to that of the in-plane
oxygen holes since there are two apical and two in-plane oxygen atoms 
per formula unit. 

An important proof for this simple model is provided by the data of the normal-state
susceptibility $\chi (T)$ (Ref.~\cite{MullerJPCM,Alex04}). The temperature 
dependence of the normal-state susceptibility in cuprates is very 
different from conventional metals.  It was found \cite{John} that 
$\chi (T)$ exhibits a broad peak at a temperature 
$T_{max}$. Such behavior was interpreted in terms of the 
2-dimensional (2D) antiferromagnetic correlation among Cu spins. 
However,  Nakano {\em et al.} \cite{Nakano} have shown 
that, at low temperatures, $\chi (T)$ strongly deviates from the 
prediction of the 2D AF model. Furthermore, the magnitude of $\chi (T)$ is 
proportional to the electronic specific heat \cite{Loram}, implying that 
the $\chi (T)$ reflects the density of states of the conduction 
electrons rather than the localized spin correlation. Alternatively, Alexandrov, 
Kabanov and Mott (AKM) explained the normal-state susceptibility data on the 
basis of their small (bi)polaron 
theory \cite{AlexPRL} which predicts a temperature-dependent spin 
susceptibility: $\chi_{AKM}(T) = 
B_{\infty}T^{-\frac{1}{2}}\exp (- \Delta_{bp} /2T)$, where $\Delta_{bp}$ is the 
bipolaron binding energy and $B_{\infty}$ is a constant depending on 
the effective masses of polarons and bipolarons \cite{AlexPRL}. If we consider  
possible coexistence with in-plane oxygen-hole polaron Fermi-liquid, the total 
susceptibility for 
the two-component system [AP (bi)polarons + in-plane oxygen-hole
polarons] is $\chi (T)$ = $f_{s}\chi_{AKM}(T)$ + 
$(1-f_{s})\chi_{F}$ + $\chi_{cV}$. Here $f_{s}$ is the 
fraction of the AP (bi)polarons, 
$\chi_{F}$ is the susceptibility contributed from the
Fermi-liquid-like polarons, and $\chi_{cV}$ is the total orbital contribution. Then
\begin{equation}\label{Te1}
\chi (T) = f_{s}B_{\infty}T^{-\frac{1}{2}}\exp (- \Delta_{bp}/2k_{B}T) + \chi_{\circ} 
\end{equation}

\begin{figure}[htb]
\vspace{-0.3cm}
    \includegraphics[height=6.5cm]{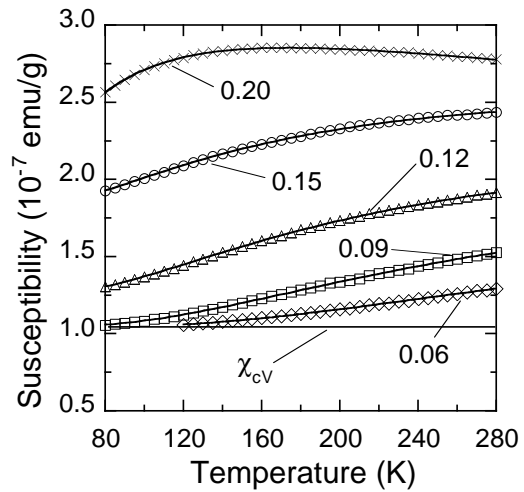}
	\vspace{0.3cm}
	\caption[~]{The temperature dependence of the normal-state 
susceptibility for La$_{1-x}$Sr$_{x}$CuO$_{4}$ (reproduced from 
Ref.~\cite{MullerJPCM}). The lowest solid line marks the total orbital 
contribution, as determined from the $^{63}$Cu Knight 
shift of La$_{1-x}$Sr$_{x}$CuO$_{4}$ (Ref. \cite{Song})}
\end{figure}

Fig.~3 shows the temperature dependence of the normal-state 
susceptibility for La$_{1-x}$Sr$_{x}$CuO$_{4}$. For $x$ = 0.20, a small Curie paramagnetic 
susceptibility has been subtracted. The lowest solid line marks the total orbital 
contribution, as determined from the $^{63}$Cu Knight 
shift of La$_{1-x}$Sr$_{x}$CuO$_{4}$ (Ref.~\cite{Song}). 
The solid lines in Fig.~3 represent the fitting curves using Eq. 
\ref{Te1}. The fitting is excellent for all the 
compositions. It is interesting to see that for $x$ $>$ 0.09, 
$(1-f_{s})\chi_{F}$ $\not =$ 0, that is, the Fermi-liquid 
type carriers are also present.  From the more extended $\chi (T)$ data 
\cite{Loram}, one sees that for $x$ $\leq$ 0.10, only (bi)polaronic 
charge carriers exist, and that the fraction of the Fermi-liquid 
carriers increases monotonically with $x$ for $x$ $>$ 0.10. Therefore, 
the theory of bipolaronic superconductivity should be applied for $x$ 
$\leq$ 0.10. For $x$ $>$ 0.1, it is possible that, due to a decrease of the charge-transfer
gap,  more in-plane Cu
holes are introduced  so
that the component of the Fermi-liquid-like carriers increases
substantially. In the overdoped region (e.g., $x$ = 0.2), the majority of
the carriers are polarons at $T_{c}$, so superconductivity should arise 
from $k$-space pairing of Fermi-liquid-like polarons.

Now we discuss how this simple model is also compatible with
ARPES results \cite{Yos} for La$_{1-x}$Sr$_{x}$CuO$_{4}$. The hole pocket centered 
around ($\pi$/2, $\pi$/2)
has been observed by ARPES for $x$ $<$ 0.18. The hole pocket should 
be associated with the mixed states of the inplane oxygen $p$
orbitals and
Cu $d_{x^{2}-y^{2}}$ orbitals.  A very flat band below the Fermi energy is seen near
($\pi$, 0) at all the doping levels. This band has been attributed to
the AP hole bipolaron state \cite{Alex96}.  Above optimal doping, the hole pocket 
around ($\pi$/2, $\pi$/2) appears to be connected
with the hole pocket around ($\pi$, 0) to form a large Fermi surface
predicted by local density approximation. Nevertheless, the measured
bare plasma energy for $x$ $>$ 0.15 is only about 1.8 eV
(Ref.~\cite{Tamasaku}), which is much smaller than the predicted value
of about 2.9 eV (Ref.~\cite{Allen}). Therefore, the apparent
connection between the two hole pockets may arise from the band-tailing
phenomenon in doped semiconductors \cite{Alex07}.

\section{The intrinsic pairing symmetry in the bulk of cuprates}

An unambiguous determination of the pairing symmetry in cuprates is 
crucial to the understanding of 
the pairing mechanism of high-temperature superconductivity. 
Many experiments have been designed to test the 
pairing symmetry in the cuprate superconductors. However, 
contradictory conclusions have been drawn from different experimental 
techniques 
\cite{Hardy,Jacobs,SFLee,Bha,Willemin,Sacuto,Kend,Wei,Review,Li,Sun,Shen,Ding,Kelley,Vob}, which 
can be classified into being bulk-sensitive and 
surface-sensitive. For example, the magnetic penetration 
depth measurements and polarized Raman scattering experiments are bulk-sensitive. 
Angle-resolved photoemission spectroscopy is in general a 
surface-sensitive technique. However, the ARPES data for 
Bi$_{2}$Sr$_{2}$CaCu$_{2}$O$_{8+y}$ should nearly reflect the bulk 
properties since the cleaved top surface contains an inactive Bi-O layer, and 
the superconducting coherent length along the 
$c$-axis is very short. The single-particle tunneling experiments along 
the CuO$_{2}$ planes can probe the bulk 
electronic density of states since the mean free path is far larger than the 
thickness of the degraded surface layer \cite{Pon}. 
In contrast, the phase-sensitive experiments based on the planar 
Josephson tunneling are rather surface sensitive, so that they might not 
probe the intrinsic bulk superconducting state if the surfaces are 
strongly degraded. Therefore, the surface- and phase-sensitive experiments do not necessarily 
provide an acid test for the intrinsic bulk gap symmetry.

Zhao \cite{ZhaoSM2001} has identified the 
intrinsic bulk pairing symmetry for hole-doped cuprates from the existing 
bulk- and nearly bulk-sensitive experimental results such as ARPES, magnetic 
penetration depth, single-particle tunneling spectra, and 
nonlinear Meissner effect. These experimental results 
consistently show that the dominant pairing symmetry in hole-doped cuprates is an extended s-wave 
with eight line nodes, that is $\Delta (\theta) = \Delta_{s} +
\Delta_{g}\cos 4\theta$, where the $g$ component $\Delta_{g}$ is
larger than the $s$-wave component $\Delta_{s}$, and $\theta$ is
measured from the Cu-O bonding direction.  Zhao
\cite{ZhaoPRB10} has also studied the 
intrinsic pairing symmetry for optimally 
electron-doped cuprates from bulk-sensitive data of Raman scattering, 
optical conductivity, magnetic penetration depth, 
directional point-contact
tunneling spectra, and nonmagnetic pair-breaking effect.  He has shown
that all these bulk-sensitive data are in 
quantitative agreement with a nearly isotropic $s$-wave gap. The
bulk-sensitive specific heat data of optimally 
electron-doped cuprates are also in quantitative agreement 
with a nearly isotropic $s$-wave gap \cite{ZhaoJPCM10}.

\begin{figure}[htb]
    \vspace{-0.3cm}
    \includegraphics[height=6.5cm]{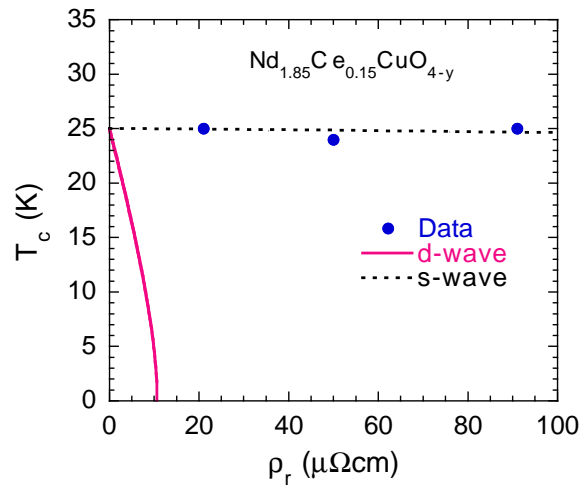}
     \vspace{-0.3cm}
 \caption[~]{$T_{c}$ as a function of residual resistivity in
optimally doped Nd$_{1.85}$Ce$_{0.15}$CuO$_{4-y}$.  The solid line is numerically 
calculated curve in terms of any $d$-wave gap
and the dotted line is the calculated curve for  
an $s$-wave gap function proportional to $(1-0.11\cos
4\theta)$. After Ref.~\cite{ZhaoPRB10}. }
\end{figure}

In particular, bulk and phase-sensitive nonmagnetic pair-breaking
effects can make clear distinction among different gap symmetries. Numerical
calculations \cite{ZhaoPRB10} for impurity pair-breaking effects indicate that
$T_{c}$ can be suppressed to zero very rapidly with both magnetic and 
nonmagnetic impurities for a $d$-wave gap: $\Delta (\theta) =
\Delta_{d}\cos 2\theta$ or a $g$-wave gap. In contrast, $T_{c}$ is
hardly suppressed with nonmagnetic impurities for nearly isotropic
$s$-wave gap \cite{ZhaoPRB10}. The fact that $T_{c}$ is
nearly independent of the residual resistivity in optimally 
electron-doped Nd$_{1.85}$Ce$_{0.15}$CuO$_{4-y}$ (see Fig.~4) should rule out any $d$-wave 
gap including a nodeless $d$-wave gap \cite{Das}.

Recent calculation based on $t-J$ model \cite{Garg} shows that the nonmagnetic
pair-breaking effect becomes less effective due to strong
electron-electron correlation. This would explain the insensitivity of
$T_{c}$ to the residual resistivity in the optimally 
electron-doped Nd$_{1.85}$Ce$_{0.15}$CuO$_{4-y}$. However,
this cannot consistently explain the very rapid $T_{c}$ suppression
with the residual resistivity in a hole-doped La$_{1.80}$Sr$_{0.20}$CuO$_{4}$
(Fig.~5b) unless the electron-electron correlation in La$_{1.80}$Sr$_{0.20}$CuO$_{4}$
is negligibly small compared with that in Nd$_{1.85}$Ce$_{0.15}$CuO$_{4-y}$.
Since the renormalized plasma 
energy $\hbar\Omega_{p}^{*}$ = 1.00 eV (see below) of La$_{1.80}$Sr$_{0.20}$CuO$_{4}$
is even much lower than that (1.64 eV) \cite{Homes}
for Nd$_{1.85}$Ce$_{0.15}$CuO$_{4-y}$, the correlation effect of the
former should be much stronger than that of the latter. If the $t-J$
model were relevant, $T_{c}$ of Nd$_{1.85}$Ce$_{0.15}$CuO$_{4-y}$
would be suppressed more rapidly than that of La$_{1.80}$Sr$_{0.20}$CuO$_{4}$.
This is in total contradiction with the experimental results (see
Figs.~4 and 5).

\begin{figure}[htb]
	 \includegraphics[height=12cm]{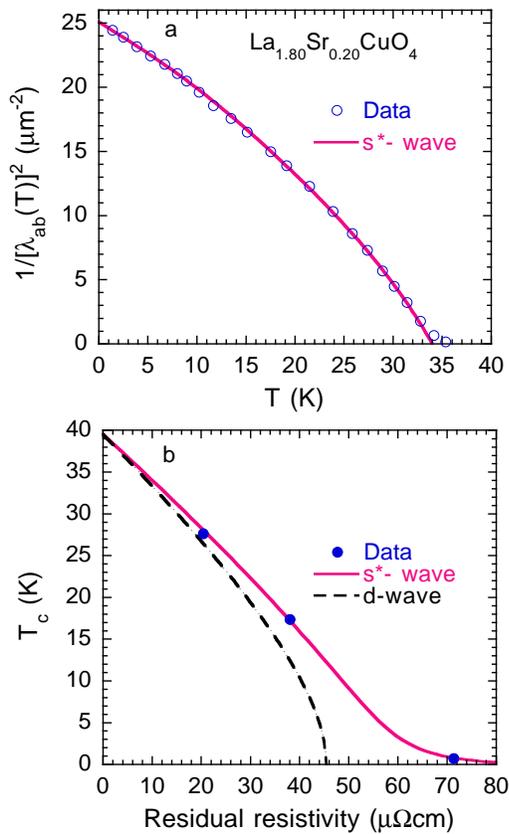}
	\caption[~]{a) Temperature dependence of the superfluid density
[$\propto$ $1/\lambda^{2}_{ab}(T)]$ for La$_{1.80}$Sr$_{0.20}$CuO$_{4}$.
The solid line is the numerically calculated curve for an extended $s$-wave gap
function: $\Delta (\theta) = 6.38 (0.24 + \cos 4\theta)$ meV. The data are
taken from Ref.~\cite{Pan}. b) $T_{c}$
dependence on the residual resistivity for La$_{1.80}$Sr$_{0.20}$CuO$_{4}$.
The solid line is calculated curve for the extended $s$-wave gap
function: $\Delta (\theta) = 6.38 (0.24 + \cos 4\theta)$ meV. The data are
taken from Ref.~\cite{Fuk}.}
\end{figure}

In fact, we can quantitatively explain the magnetic penetration depth 
and non-magnetic pair breaking effect in La$_{1.80}$Sr$_{0.20}$CuO$_{4}$ even without taking into account electron-electron correlation.
Fig.~5a shows the temperature dependence of the superfluid density
[$\propto$ $1/\lambda^{2}_{ab}(T)]$ for La$_{1.80}$Sr$_{0.20}$CuO$_{4}$.
The solid line is the numerically calculated curve for an extended $s$-wave ($s^{*}$-wave) gap
function: $\Delta (\theta) = 6.38 (0.24 + \cos 4\theta)$ meV using the
standard formula \cite{ZhaoSM2001} and the BCS temperature dependence of the gap. This
gap function has a similar form as that for YBa$_{2}$Cu$_{3}$O$_{7}$
with a similar doping \cite{ZhaoSM2001}. Fig.~5b shows $T_{c}$
dependence on the residual resistivity for La$_{1.80}$Sr$_{0.20}$CuO$_{4}$.
The solid line is calculated curve for the extended $s$-wave gap
function: $\Delta (\theta) = 6.38 (0.24 + \cos 4\theta)$ meV obtained from
the penetration depth data (Fig.~5a) using the standard formula
\cite{ZhaoSM2001} and the renormalized plasma 
energy $\hbar\Omega_{p}^{*}$ = 1.00 eV that is calculated from the
measured zero-temperature penetration depth (2000~\AA) (see Fig.~5a). It
is apparent that the calculated curve is in quantitative agreement
with the data. In contrast, the calculated curve (dashed line) for
$d$-wave gap is significantly off from the data. 

When hole doping is reduced, the $s$-wave component increases so that 
the nonmagnetic pair-breaking effect becomes weaker. Since the
bipolaronic charge carriers become dominant in underdoped cuprates,
nonmagnetic impurities do not break the pairs, but can still suppress
$T_{c}$ via localization effect. 

With the two-carrier scenario in hole-doped cuprates, we can
consistently explain surface and phase-sensitive experiments which probe 
$d$-wave order-parameter symmetry at surfaces and interfaces. 
Experiments on
hole-doped cuprates \cite{Bet,Mann} have shown that surfaces
and interfaces are significantly underdoped so that bipolaronic
charge carriers are dominant over polaronic Fermi-liquid-like ones.  Since the 
order-parameter symmetry
of Bose-Einstein condensate of bipolarons are $d$-wave
\cite{AlexSM98}, the
surface and phase-sensitive experiments should probe the dominant $d$-wave
order-parameter symmetry, in agreement with experiments \cite{Review}.  For electron-doped ($n$-type) cuprates, the penetration depth 
\cite{Alff,Kim} and point-contact tunneling spectra \cite{Kas,Bis,Qaz,Shan05,Shan08}
consistently suggest that the gap symmetry in deeply underdoped
samples should be $d$-wave and change to a nodeless $s$-wave when the
doping level is above a critical value, in agreement with the
theoretical prediction based on a phonon-mediated pairing mechanism
\cite{AlexSM08}. This scenario can also explain
the $d$-wave gap symmetry inferred from 
surface-sensitive experiments if surfaces or
interfaces are deeply underdoped.  Another possibility is that superconductivity in deeply underdoped $n$-type cuprates is also due to the Bose-Einstein condensation of local pairs. In this case, the superconducting quasi-particle gap is found to have a $d$-wave symmetry \cite{Wei}.   In sharp contrast to the standard theories of 
Bogoliubov quasi-particle excitations, the quasi-particle  gap is shown \cite{Wei} to originate from anomalous kinetic process, completely unrelated to the pairing symmetry.  
 
\section{Concluding remarks}

We have presented some crucial
experiments that place essential constraints on the pairing mechanism 
of high-temperature superconductivity. The observed unconventional oxygen-isotope 
effects in cuprates have clearly shown strong electron-phonon interactions and the 
existence of polarons and/or bipolarons.  Angle-resolved photoemission and tunneling spectra
have provided direct evidence for strong coupling to multiple-phonon modes.
In contrast, these spectra do not show strong coupling features expected for magnetic 
resonance modes. Angle-resolved photoemission spectra
and the oxygen-isotope effect on the antiferromagnetic exchange
energy $J$ in undoped parent compounds consistently show that the
polaron binding energy is about 2 eV, which is over one order of magnitude larger than 
$J$ = 0.14 eV.  In addition, a photoinduced lattice expansion experiment  \cite{BozPRB} and various optical data \cite{Mak} also provide evidence for strong electron-phonon coupling.

The normal-state spin-susceptibility data of hole-doped cuprates
indicate that intersite bipolarons are the dominant charge carriers in the 
underdoped region while the component of Fermi-liquid-like polarons is dominant 
in the overdoped region. This  two component scenario is also supported by the femtosecond optical spectroscopic experiment on 
(Y$_{1-x}$Ca$_{x}$)Ba$_{2}$Cu$_{3}$O$_{7-y}$.    Various bulk-sensitive experiments  consistently
demonstrate that the intrinsic gap (pairing) symmetry for the Fermi-liquid-like
component is anisotropic $s$-wave  while  surface- and phase-sensitive experiments see $d$-wave order-parameter symmetry at intrinsically underdoped surfaces and interfaces where the dominant bipolaronic charge carriers are condensed into a $d$-wave superconducting state \cite{AlexSM98,Wei}.

~\\
$^{*}$ gzhao2@calstatela.edu

\bibliographystyle{prsty}

\end{document}